\documentclass[conference]{IEEEtran}
\IEEEoverridecommandlockouts
\usepackage{cite}
\usepackage{amsmath,amssymb,amsfonts}
\usepackage{algorithmic}
\usepackage{graphicx}
\usepackage{textcomp}
\usepackage{xcolor}
\usepackage{multirow}
\def\BibTeX{{\rm B\kern-.05em{\sc i\kern-.025em b}\kern-.08em
    T\kern-.1667em\lower.7ex\hbox{E}\kern-.125emX}}
\begin{document}

\title{Hand-drawn Symbol Recognition of Surgical Flowsheet Graphs with Deep Image Segmentation}

\author{\IEEEauthorblockN{William Adorno III}
\IEEEauthorblockA{\textit{Department of Engineering} \\
\textit{Systems and Environment} \\
\textit{University of Virginia}\\
Charlottesville, VA USA \\
wa3mr@virginia.edu}
\and
\IEEEauthorblockN{Angela Yi}
\IEEEauthorblockA{\textit{Department of Engineering} \\
\textit{Systems and Environment} \\
\textit{University of VIrginia }\\
Charlottesville, VA USA \\
ay2ug@virginia.edu}
\and
\IEEEauthorblockN{Marcel Durieux}
\IEEEauthorblockA{\textit{Department of Anesthesiology} \\
\textit{School of Medicine} \\
\textit{University of Virginia}\\
Charlottesville, VA USA \\
durieux@virginia.edu}
\and 
\IEEEauthorblockN{Donald Brown}
\IEEEauthorblockA{
\textit{School of Data Science} \\
\textit{University of Virginia}\\
Charlottesville, VA USA \\
deb@virginia.edu}
}

\maketitle

\begin{abstract}
Perioperative data are essential to investigating the causes of adverse surgical outcomes. In some low to middle income countries, these data are computationally inaccessible due to a lack of digitization of surgical flowsheets. In this paper, we present a deep image segmentation approach using a U-Net architecture that can detect hand-drawn symbols on a flowsheet graph. The segmentation mask outputs are post-processed with techniques unique to each symbol to convert into numeric values. The U-Net method can detect, at the appropriate time intervals, the symbols for heart rate and blood pressure with over 99 percent accuracy. Over 95 percent of the predictions fall within an absolute error of five when compared to the actual value. The deep learning model outperformed template matching even with a small size of annotated images available for the training set. 

\end{abstract}


\section{Introduction}
Many hospitals in developing countries still use paper flowsheets to collect patient data before, during, and after surgery. The data are recorded by hand, which makes it difficult to digitize and conduct further research. The paper flowsheets contain crucial data that can be analyzed to assess areas such as the rate and cause of perioperative mortality~\cite{rickard2016associations}. Digitization of the perioperative data can enable follow-on research to reduce perioperative mortality rate and improve surgical outcomes. As an example of the problem, the University Teaching Hospital of Kigali in Rwanda has six years' worth of surgical flowsheets that have not been analyzed to understand adverse surgical outcomes. Recently, researchers at the University of Virginia developed a system to scan the flowsheets and extract data from important sections \cite{9106679}. This paper builds on that effort to solve the problem of reading hand-drawn symbols in graphs that represent values over time for heart rate and blood pressure.

Our approach performs deep segmentation with a U-Net architecture that predicts the locations of symbols within segmentation masks. This deep learning method is trained and tested with 29 annotated graph images. The segmentation masks are then post-processed with symbol-specific techniques that convert the masks into numerical values. We measure the effectiveness of digitization by using detection accuracy, residual errors, bias, and variance. An additional 32 test images, with true values recorded for heart rate and blood pressures, are used in this evaluation. The U-Net method results show greater performance when compared to template matching, a popular image processing approach. Importantly, our deep image segmentation method demonstrated improved results even with a small size of annotated images used for model training. In the remainder of this paper, we discuss the related work in symbol recognition, the methodology applied to develop the U-Net model and post-process the segmentation masks, and the results of the comparison between the two approaches.

\section{Approaches to Symbol Recognition}

\subsection{Hand-drawn Symbol Recognition}

Hand-drawn symbol recognition is a topic within the computer vision field that has had intensive research during the past four decades. Previous work has looked at recognizing hand-written text~\cite{krishnan2016deep}, hand drawn chemical organic ring structure symbols~\cite{zheng2019recognition}, electrical engineering drawings~\cite{moreno2019new}, and more. The MNIST database of handwritten digits is one of the most popular benchmarks datasets for deep learning classification problems \cite{deng2012mnist}. While these applications had success with image classification, they require large training sets and images of the same size. As a result, the successful machine learning classification approaches for these areas do not extend to finding the locations of small symbols within a scanned document. 

\subsection{Template Matching}

The lack of effective machine learning techniques for symbol recognition from scanned documents has led the use of image processing techniques \cite{fornes2010rotation, kara2005image}. However, more promising results have been offered by template matching techniques that scan the image and identify the location of the best matches \cite{brunelli2009template}. Template matching is widely used in image recognition and computer vision applications including medical image processing~\cite{adagale2013image}, automatic car driving~\cite{andrey2006automatic}, and defect inspection systems~\cite{yoon2008effective}. Template matching involves scanning an image with a smaller image template to search for areas with high cross-correlation ~\cite{choi2002novel}. The template image contains the object or symbol that is desired to find within the source image. The main hyperparameter of template matching is the cross-correlation threshold to determine if there is a confident template match. When the threshold is too high, the rate of false negative increases, whereas a low threshold increases the rate of false positives.

Some difficulties with template matching are accounting for variations in shift, rotation. and scale of objects~\cite{hashemi2016template} and high computational costs. There is research in developing techniques that are invariant to these factors \cite{kim2007grayscale}. For the problem of symbol recognition in scanned surgical flowsheets, variations in rotation and scale are not a concern. The symbols have nearly fixed rotations and are typically a similar size. 

\subsection{Convolutional Neural Networks}

Classical feed-forward neural networks are ill-suited for image data due to the immense number of nodes required to fully-connect with the input image. Convolutional Neural Networks (CNN) correct these problems and have become the leading deep learning approach for modeling image data.  The inclusion of convolutional and pooling layers are unique aspects of CNNs that differ from feed-forward neural networks \cite{goodfellow2016deep}. Each convolutional layer consists of multiple filters known as kernels. The kernels are scanned through the image to detect important features at certain locations resulting in a feature map. Pooling layers are applied on the feature maps to reduce the amount of stored data, while also making the model invariant to small spatial changes in the input image \cite{scherer2010evaluation}. 

Series of convolutional and pooling layers are stacked together to form convolutional blocks. For classification and regression problems, the input image is processed through  one or more convolutional blocks to create a vector representation of image features. This vector is then processed through fully-connected layers to complete the prediction process \cite{goodfellow2016deep}. The weights and biases of the model, which includes the kernels, are automatically optimized during training using backpropagation and stochastic gradient descent \cite{lecun2015deep}. Researchers have developed CNN architecture variations to improve prediction performance such as incorporating skip connections into convolution blocks to create residual blocks. Residual blocks allow CNNs to become much deeper without degradation by enabling activations to skip layers when necessary \cite{he2016deep}.

\subsection{Deep Image Segmentation}

CNN models for image segmentation differ from traditional CNNs, because the output is now a segmentation mask that is the same size as the input image. Each pixel of the segmentation mask contains a classification prediction. Some examples of applied image segmentation include: extracting cellular features from medical biopsy images \cite{falk2019u}, detecting cracks in structures for civil engineers \cite{alipour2020increasing}, and identifying surroundings for self-driving cars \cite{treml2016speeding}. The U-Net image segmentation approach was originally developed for the biomedical domain with the goal of not requiring large training sets \cite{ronneberger2015u}. 

The U-Net architecture mimics an encoder-decoder network where the image is contracted and then expanded back to the original size. Each convolutional block or down-sampling step includes two convolutional layers and then a max-pooling layer. The original U-Net model contains four convolutional blocks until the bottom of the ``U-shape'' is reached \cite{ronneberger2015u}. Next, there are four sets of deconvolutional blocks or up-sampling steps to restore the size of the image. This architecture is considered fully-convolutional, because it does not include any feed-forward layers. To further assist with reconstructing the image during up-sampling, U-Nets also concatenate high-resolution features from the contracting side with the expanding side of the ``U''. 

Researchers have also developed adaptations to the original U-Net architecture. The Residual U-Net incorporates residual blocks into the architecture to increase deepness of the model, while still propagating information quickly \cite{zhang2018road}. The loss function used for training U-Net models is typically an Intersection-over-Union (IoU) technique. These techniques tend to outperform cross entropy when the segmentation is sparse or a small fraction of the total image. One of the most popular IoU techniques is the Dice similarity coefficient \cite{dice1945measures}. Dice Loss (DL) was incorporated into a CNN architecture by \cite{milletari2016v} and is defined as:

\begin{equation}\label{eq:dice_loss}
    DL =  1 - \frac{2 \sum{i}^{N} p_i g_i}{\sum{i}^{N} p_i^2 + \sum{i}^{N} g_i^2}
\end{equation}

where $p_i$ is the predicted binary segmentation volume, $g_i$ is the ground truth binary volume, and $N$ is the total number of images. 

\section{Methodology}

This section explains the process of annotating image data, training a U-Net image segmentation model, and converting the predictions into numerical time series values. We also explain the methods used in the template matching baseline approach. 

\subsection{Annotations and Segmentation Masks}

Deep image segmentation is a supervised approach that requires annotated masks to train the model and represent ground truth. The true locations of the hand-drawn symbols on the graph must be annotated to build the training, validation, and test sets. A total of 29 images were annotated for all three symbols: heart rate, diastolic blood pressure (BP), and systolic BP. During surgery, heart rate values are represented by drawing small circles on the graph. Heart rates are captured by annotating the locations of the center of the circles. The two blood pressures are represented as arrows on the graph with the diastolic BP pointed upward and systolic BP pointed downward. The BP arrows are captured by annotating a rectangle that surrounds the symbol. 

\begin{figure}
\center
\includegraphics[width=0.4\textwidth]{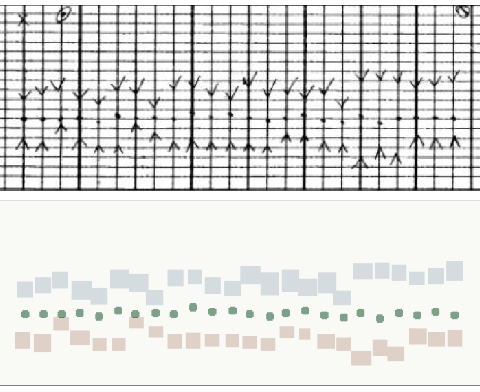} 
\caption{Image annotation example. The top image is one of the training set images. The bottom image contains an overlay of the three ground truth segmentation masks. Systolic blood pressures are in blue, the heart rates are in green, and diastolic blood pressures are in red.}
\label{fig-annotation}
\end{figure}

The annotations for each symbol are processed into separate ground truth masks. Three different U-Net models are trained for each of the three symbols since the segmentation output may overlap. The BP rectangle annotations remain the same on ground truth masks. The heart rate masks become larger circles with a radius of three pixels centered at the annotation coordinate. Fig. \ref{fig-annotation} shows an example of the three segmentation masks overlayed onto the same image. The graph images are very rectangular with an image size of $164 \times 990$ pixels. Input images for U-Nets are typically square with dimensions that are a power of two. To conform to this structure, we chose to zero-pad the graphs to create $1024 \times 1024$ pixel images in order to fit and train on the entire graph image.

\subsection{Training the U-Net Model}

The 29 annotated images are allocated to training, validation, and test sets with sizes of 23, 4, and 2, respectively. The validation set is utilized during training to prevent over-fitting. The validation loss is evaluated after each epoch and only the best performing model parameters are saved. A Residual U-Net model similar to the approach implemented by \cite{zhang2018road} is trained for each of the three symbol types. This architecture contains four residual blocks. Each residual block contains two convolutional blocks and a skip connection that allows for the entire residual unit to be bypassed. Each convolutional block contains batch normalization, a convolutional layer, and a Rectified Linear Unit (ReLU) activation function \cite{ioffe2015batch, dahl2013improving}. The first residual block contains eight filters in the convolutional layer and then this number doubles in each of the following residual blocks until there are 128 filters at the bridge between down-sampling and up-sampling. All filters have a kernel size of $[3, 3]$. The architecture described contains $1,185,405$ total parameters.  

\begin{figure}
\center
\includegraphics[width=0.42\textwidth]{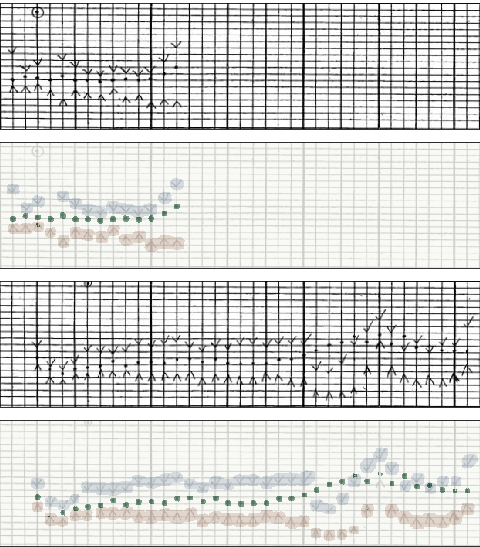} 
\caption{Test set image prediction. The first and third images are test set image inputs. The second and fourth images have the three prediction masks overlayed. Systolic blood pressures are in blue, the heart rates are in green, and diastolic blood pressures are in red.}
\label{fig-test}
\end{figure}

Each Residual U-Net was trained for 200 epochs and with a constant learning rate of $2 \times 10^{-3}$ to optimize Dice loss. After the model is trained, the prediction segmentation masks are created using a probability threshold of $0.50$. The two test set images are processed through the three models to evaluate performance. The average Dice coefficient scores on the test set are $0.832$, $0.882$, $0.878$ for heart rate, diastolic BP, and systolic BP, respectively. Dice scores this high usually indicate a successful image segmentation. Fig. \ref{fig-test} shows the prediction results for the two test set images. The prediction masks tend to detect arrows with a circle-like shape rather than a complete square. This may cause a slight decrease in the Dice score for the BPs, because the corners of the squares are not detected. It is not surprising that the BPs are easier to detect than heart rates. Some of the heart rate values are drawn faintly and can blend in with the graph's gridlines.

While the segmentation results on the test set are promising, we should confirm that these results are valid on more than just this small test set. To further test the performance of this approach, a 10-fold cross validation was performed on all three models \cite{fushiki2011estimation}. Since there are total of 29 images, the first nine folds will contain three images in the test set and the tenth fold contains two images. Table \ref{tab-crossval} shows summary statistics for the 10-fold cross validation. The performance of the U-Net model was consistent over all ten folds. The minimum values of $0.724$, $0.811$, and $0.801$ for heart rate, diastolic BP, and systolic BP, respectively, would be considered adequate in many image segmentation applications. 

\begin{table}[htbp]
\caption{10-Fold Cross Validation Results on U-Net Model}
\begin{center}
\begin{tabular}{|c|c|c|c|}
\hline
\textbf{Summary}&\multicolumn{3}{|c|}{\textbf{Test Set Dice Coefficient}} \\
\cline{2-4} 
\textbf{Statistic} & \textbf{\textit{Heart Rate}}& \textbf{\textit{Diastolic BP}}& \textbf{\textit{Systolic BP}} \\
\hline
Median & $0.800$ & $0.847$ & $0.843$  \\
\hline
Average & $0.789$ & $0.844$ & $0.845$  \\
\hline
Standard Dev. & $0.037$ & $0.015$ & $0.029$  \\
\hline
Minimum & $0.724$ & $0.811$ & $0.801$  \\
\hline
Maximum & $0.831$ & $0.860$ & $0.887$  \\
\hline
\multicolumn{4}{l}{}
\end{tabular}
\label{tab-crossval}
\end{center}
\end{table}

\subsection{Post-Processing of Segmentation Masks}

The main objective of this research is to convert the hand-drawn symbols into numerical values to populate time series data. The output of the U-Net models are three segmentation masks that describe the location of all detected heart rate and BP symbols. A post-processing algorithm is required to convert segmentation masks into numerical values. The heart rate masks differ from the BP masks, so each has a unique post-processing technique. The horizontal axis of the graph is the time axis with each gridline representing a 5-minute increment. The average pixel distance in the 5-minute intervals is approximately 16.5 pixels. This makes certain aspects of post-processing difficult since pixels can only be represented as integers. The result after each conversion is a two-column matrix where the first column contains the time values and the second column contains the values for heart rates and BPs that correspond to each time. 

The main steps for each post-processing method are listed and then followed by a more detailed explanation. The post-processing technique to convert each heart rate segmentation mask prediction into numerical time series values is as follows:

\begin{enumerate}
	\item Remove objects that are less than 12 contiguous pixels
	\item Perform an opening morphology with a disk structuring element (radius $=2$)
	\item Process cleaned mask through \textit{RegionProps} function from the Python Skimage package
	\item Time value = the $y$-coordinate of each object centroid from \textit{RegionProps} divided by 16.5 pixels
	\item Heart rate value = image height minus $x$-coordinate of each object centroid
	\item Sort all values by the time value column
	\item Convert values from pixel units into the proper graph scale with (\ref{eq:pixel_bpm})
\end{enumerate}

The first step removes false detections that are much smaller than the typical heart rate symbol size. The opening morphology takes this a step further and ensures that the remaining objects are at least the size of circle with a radius of 2 pixels. The \textit{RegionProps} function generates a large amount of information regarding all of the contiguous objects in the cleaned segmentation mask. One of the data points captured by this function is the centroid of each found object. This centroid is critical in generating the proper time and heart rate values. The value is subtracted from the image height, because the order of the $y$-axis is reversed between the image and the actual graph. Finally, the results are still in pixel units, so they must be converted into the proper scale of the graph. 

The graph's $y$-axis ranges from 0 to 210 Beats Per Minute (BPM) and the height of the original image is 164 pixels. All of the gridlines of the graph are equidistant (10 BPM) except for the bottom row which ranges from 0 to 30. Therefore, the bottom row is factored out before doing the slope conversion. The pixel height of the bottom row is approximately 13 pixels. Equation \ref{eq:pixel_bpm} is applied to convert the pixel values into BPM:

\begin{equation}\label{eq:pixel_bpm}
    V_t = \frac{p_t - 13}{h - 13} * (210 - 30) + 30
\end{equation}

where $V_t$ is the converted value in the appropriate units at time $t$, $p_t$ is the pixel value at time $t$, and $h$ is the height of the image. The end-user of this data prefers that it is stored in an integer format, so all $V_t$ are rounded to the nearest integer.

The post-processing method for BP symbols is quite different from the heart rate approach, because the BP segmentation masks contain much larger objects that are often connected together. Instead of utilizing \textit{RegionProps} to capture objects, the BP values can be obtained by iterating through the image at certain pixel intervals. The post-processing technique to convert BP segmentation masks into millimeters of mercury (mmHg) units is as follows:

\begin{enumerate}
	\item Remove objects that are less than 12 contiguous pixels
	\item Iterate through the image from left to right using a pixel interval that alternates between $+16$ and $+17$
	\item At each interval, iterate from the bottom of the image to top until the systolic BP value changes from $0$ to $1$
	\item Systolic BP value = image height minus the pixel value where iteration stopped
	\item At each interval, iterate from the top of the image to bottom until the diastolic BP value changes from $0$ to $1$
	\item Diastolic BP value = image height minus the pixel value where iteration stopped
	\item Convert values from pixel units into the proper graph scale using (\ref{eq:pixel_bpm}) plus an additional correction factor
\end{enumerate}

The pixel interval alternates between $+16$ and $+17$ to avoid drifting and sampling at the wrong time. The heart rate symbols are very small, so some of these objects would be missed with an iteration approach. The BP symbols are wide enough to be captured in this manner. The predicted BP segmentation masks appear to surround the arrow symbol. The actual value of each BP is the tip of the arrow, so there is some bias created when only iterating to the point at which the mask first appears. If this bias can be estimated, we can add a correction factor to (\ref{eq:pixel_bpm}) that accounts for this buffer region between the mask and the tip of the arrow. At this point in the methodology development, we do not have an estimate for the correction factor, but we will re-address this issue in the Results section when we calculate prediction errors. Lastly, to avoid false detections long after the surgery has ended, the post-processing algorithm will stop looking for BP values when no values are detected for ten consecutive samples.  

\subsection{Template Matching Method}
\begin{figure}
\center
\includegraphics[width=0.35\textwidth]{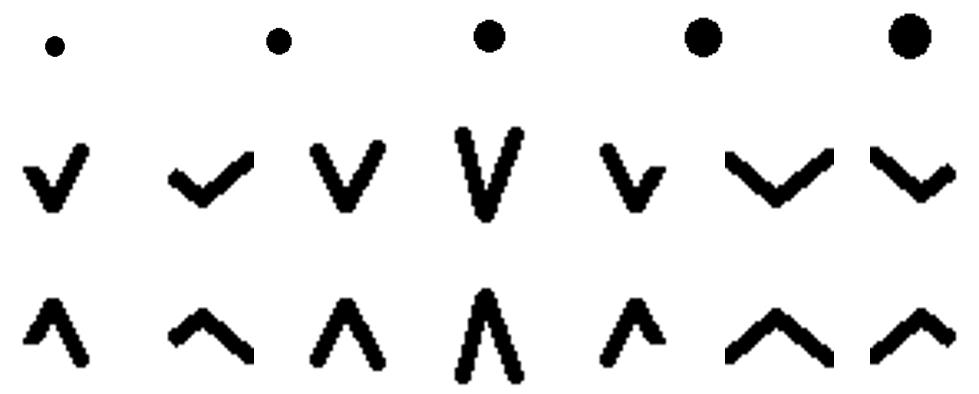} 
\caption{Templates Used. The first row shows the different sizes of circles used as templates for heart rates. The second and third row are templates used for systolic blood pressure and diastolic blood pressure, respectively. }
\label{fig-templates}
\end{figure}

Due to the variability in symbol drawings, template matching for graph digitization was performed with multiple templates for each symbol. The size of circle is the main source of variability that can be modeled for heart rates. Each template has an individual cross-correlation threshold. The thresholds are fairly similar within each symbol. Templates with five different size circles are used to detect heart rates and have an average match threshold of 0.13. For each BP, seven different templates are utilized that contain arrows of different angles and sizes. The average cross-correlation thresholds used for diastolic BP and systolic BP templates were 0.16 and 0.18, respectively. Increasing the number of templates can improve prediction performance, but eventually it becomes difficult to manually design new alternatives and not exceed computational costs or time. The templates used are shown in Fig. \ref{fig-templates}. Cross-correlation is used as the metric to evaluate a template match. If the cross-correlation exceeds a certain threshold, it is considered a positive detection of a symbol. The selected thresholds produced the lowest mean squared error on the same training set used for the U-Net model.

Template matching produces a list of coordinates that represent the center pixel of each positive match. Post-processing is performed to eliminate the excessive matches and to convert location coordinates into numerical values of the measurements. To eliminate excessive matches, the algorithm only searches for matches that occur at the 5-minute time intervals. At each given time, the algorithm iterates from the top of the image to the bottom and finds the location of the match. Once a match is found, all points within 16 pixels on the $x$-axis are eliminated. This removes most of the false positive matches. The conversion of values from pixel units into the proper graph scale was also performed by using (\ref{eq:pixel_bpm}).

\section{Results}

We next evaluate how well the entire methodology is able to detect symbols at the right time and within the tolerance range of the true value. To perform this evaluation, a completely new test set of 32 images was obtained. This test set includes the true value of all three symbols recorded at each time interval. The evaluation is a two-stage process. First, we assess the ability to detect a symbol at each appropriate time interval. Second, we determine how accurate the detected symbols are to the truth values. The performance of the U-Net approach is compared to the Template Matching (TM) baseline.

\subsection{Symbol Detection Performance}

Each graph contains enough space along the $x$-axis to record 5 hours of surgery data. On every image, we store 59 time series values for each symbol to cover most of this 5-hour period, although surgeries are often between 1 to 3 hours long. If there is no value recorded at a particular time, the time series entry remains blank. The 32 test images contain a total of 1,888 time series entries. Of these entries, 654, 659, and 657 contain numerical values for heart rate, diastolic BP, and systolic BP, respectively. The remaining entries are blank, because they occur outside the surgery time frame or a symbol was not recorded at that 5-minute interval. For the terminology of the detection performance analysis, a ``positive'' is when the a true value exists for a symbol and a ``negative'' is when the time entry is blank. A true positive occurs if the model detects a symbol when a true value exists. A true negative occurs when both the model and the true value are blank. A false positive occurs if the models detects a symbol when the true is actually blank, while a false negative occurs if the model does not detect a symbol when a true value exists.

In Table \ref{tab-detect_comp}, the results for precision, recall, and F$_1$ score are displayed for each symbol and method \cite{sokolova2009systematic}. For all three symbols, the U-Net model outperforms the Template Matching (TM) baseline in all three detection metrics. The U-Net approach rarely produced false positives or false negatives, while the TM approach produced a large number of false negatives. While U-Net outperformed TM in all metrics, the results are accurate for both methods and some of the differences are small. 

\begin{table}[htbp]
\caption{Symbol Detection Performance Comparison}
\begin{center}
\begin{tabular}{|c|c|c|c|c|}
\hline
\textbf{Symbol} & \textbf{Model}& \textbf{\textit{Precision}}& \textbf{\textit{Recall}}& \textbf{\textit{F$_1$ Score}} \\
\hline
\multirow{2}{1.5cm}{Heart Rate} & U-Net & 0.997 & 0.976 & 0.987 \\
\cline{2-5}
 & TM & 0.965 & 0.873 & 0.936\\
\hline
\multirow{2}{1.5cm}{Diastolic BP} & U-Net & 0.997 & 0.989 & 0.994  \\
\cline{2-5}
 & TM & 0.954 & 0.822 & 0.911 \\
\hline
\multirow{2}{1.5cm}{Systolic BP} & U-Net & 0.986 & 0.995 & 0.998 \\
\cline{2-5}
 & TM & 0.969 & 0.860 & 0.930 \\
\hline
\end{tabular}
\label{tab-detect_comp}
\end{center}
\end{table}

 To perform a statistical comparison of detection rates between U-Net and TM, the 32 images are treated as independent samples. For each image, there is a performance metric sample for both U-Net and TM. We want to test whether the difference between U-Net and TM performance is significantly greater than zero. The null hypothesis of this test is $H_0: \mu_d = 0$ where $\mu_d$ is the true mean difference. The upper-tailed alternative hypothesis is $H_A: \mu_d > 0$ to test if the U-Net results are significantly greater than TM. The $d_i$ of each image, the difference between U-Net and TM, appears to be approximately normally distributed for all symbols and performance metrics, so a $t$-test is appropriate on this data. The test statistic is $t = \frac{\bar{d}-\mu_d}{\hat{\sigma}/\sqrt{n}}$ where $\bar{d}$ is the sample mean difference, $\hat{\sigma}$ is the sample standard deviation, and $n$ is the total number of images. The denominator $\hat{\sigma}/\sqrt{n}$ is known as the standard error. The test statistic is compared to the appropriate $t$-critical value for a 95\% confidence to determine the result of the hypothesis test. 

Table \ref{tab-detect_hr} shows the hypothesis test results for the heart rate symbol. For all three performance metrics, the U-Net method has a positive sample mean difference over TM. At  $\alpha = 0.05$, all three of these mean differences were large enough to reject the null hypothesis that $\mu_d = 0$. Therefore, this shows statistical evidence that the U-Net method significantly outperforms TM in detecting heart rate symbols.

\begin{table}[htbp]
\caption{Heart Rate Detection Statistical Comparison}
\begin{center}
\begin{tabular}{|c|c|c|c|}
\hline
\textbf{$t$-test parameters} & \textbf{\textit{Precision}}& \textbf{\textit{Recall}}& \textbf{\textit{F$_1$ Score}} \\
\hline
Mean Difference & 0.029 & 0.104 & 0.067 \\
\hline
Standard Error & 0.0071 & 0.0128 & 0.0075 \\
\hline
$t$-statistic & 4.13 & 8.11 & 8.82 \\
\hline
$p$-value & $1.3 \times 10^{-4}$ & $1.8 \times 10^{-9}$ & $2.9 \times 10^{-10}$ \\
\hline
Result ($\alpha = 0.05$) & Reject $H_0$ & Reject $H_0$ & Reject $H_0$ \\
\hline
\end{tabular}
\label{tab-detect_hr}
\end{center}
\end{table}

Table \ref{tab-detect_dbp} shows the hypothesis test results for the diastolic BP symbol. For all three performance metrics, the U-Net method has a positive sample mean difference over TM. At  $\alpha = 0.05$, all three of these mean differences were large enough to reject the null hypothesis that $\mu_d = 0$. Therefore, this shows statistical evidence that the U-Net method also significantly outperforms TM in detecting diastolic BP symbols.

\begin{table}[htbp]
\caption{Diastolic BP Detection Statistical Comparison}
\begin{center}
\begin{tabular}{|c|c|c|c|}
\hline
\textbf{$t$-test parameters} & \textbf{\textit{Precision}}& \textbf{\textit{Recall}}& \textbf{\textit{F$_1$ Score}} \\
\hline
Mean Difference & 0.040 & 0.170 & 0.105 \\
\hline
Standard Error & 0.0088 & 0.0125 & 0.0054 \\
\hline
$t$-statistic & 4.59 & 13.56 & 19.33 \\
\hline
$p$-value & $3.4 \times 10^{-5}$ & $7.1 \times 10^{-15}$ & $3.8 \times 10^{-19}$ \\
\hline
Result ($\alpha = 0.05$) & Reject $H_0$ & Reject $H_0$ & Reject $H_0$ \\
\hline
\end{tabular}
\label{tab-detect_dbp}
\end{center}
\end{table}

Table \ref{tab-detect_sbp} shows the hypothesis test results for the systolic BP symbol. For all three performance metrics, the U-Net method has a positive sample mean difference over TM. At  $\alpha = 0.05$, the mean differences for recall and F$_1$ score were large enough to reject the null hypothesis that $\mu_d = 0$. The $p$-value for precision is slightly above the $\alpha$ value of 0.05, so the null hypothesis failed to reject. The U-Net approach produced less false positives for SBP than TM, but not few enough to conclude significance. Some of the false positives are caused by occasional markings on images to show the surgery has ended. These markings include an area that resembles a downward pointed arrow. 

\begin{table}[htbp]
\caption{Systolic BP Detection Statistical Comparison}
\begin{center}
\begin{tabular}{|c|c|c|c|}
\hline
\textbf{$t$-test parameters} & \textbf{\textit{Precision}}& \textbf{\textit{Recall}}& \textbf{\textit{F$_1$ Score}} \\
\hline
Mean Difference & 0.015 & 0.137 & 0.076 \\
\hline
Standard Error & 0.0091 & 0.0114 & 0.0087 \\
\hline
$t$-statistic & 1.65 & 12.02 & 8.73 \\
\hline
$p$-value & $0.054$ & $1.7 \times 10^{-13}$ & $3.7 \times 10^{-10}$ \\
\hline
Result ($\alpha = 0.05$) & FTR* $H_0$ & Reject $H_0$ & Reject $H_0$ \\
\hline
\multicolumn{4}{l}{$^{\mathrm{*}}$FTR - Fail to Reject}
\end{tabular}
\label{tab-detect_sbp}
\end{center}
\end{table}

\subsection{Prediction Errors}

The performance evaluation of these methods is a two-stage process. First, the detection rate assessment showed how well the two approaches detected symbols at the proper times, while also avoiding false detections. Next, we must evaluate if the detections are closely accurate with the ground truth numerical value for all three symbols. The prediction errors can only be evaluated where ground truth value exist, so false positives are excluded from this section of the results. False negatives do not have predicted numerical values, so we first concentrate on only the error amongst the true positives.

Table \ref{tab-true_pos} shows the results for true positive errors across the entire dataset of 32 images. The true positives varied for each method, so a paired statistical comparison is not possible here. The mean errors for heart rate were both near zero which indicates minimal bias. All values are ultimately stored as integers, so slight biases are not impactful. The mean errors for BPs do reveal some biases that we did expect to find. The U-Net BP mean errors are both off by about $4.0$ mmHg. This is due to the segmentation masks having a buffer between the edge of the each object and the tip of the arrow. To correct this bias for the U-Net method, a $-4$ correction factor is added to (\ref{eq:pixel_bpm}) for diastolic BP and a $+4$ correction factor is added to (\ref{eq:pixel_bpm}) for systolic BP. The TM mean error for BP also has a bias of $+2.60$. This bias is due to selecting the best template matches by iterating from the top of the image to the bottom. This can be corrected by subtracting $2.60$ from each BP value conversion for the TM method. The sample standard deviations were much lower for the U-Net method which indicates a more accurate prediction.

\begin{table}[htbp]
\caption{True Positive Error Results}
\begin{center}
\begin{tabular}{|c|c|c|c|c|}
\hline
\textbf{Method} & \textbf{Statistic} & \textbf{\textit{Heart Rate}}& \textbf{\textit{Diastolic BP}}& \textbf{\textit{Systolic BP}} \\
\hline
\multirow{3}{8mm}{U-Net} & $n$ & 638 & 652 & 654 \\
\cline{2-5}
& Mean Error & -0.50 & 4.01 & -3.96 \\
\cline{2-5}
& SD* & 2.11 & 2.46 & 2.02 \\
\hline
\multirow{3}{8mm}{TM} & $n$ & 571 & 542 & 565 \\
\cline{2-5}
& Mean Error & 0.72 & 2.60 & 2.62 \\
\cline{2-5}
& SD* & 4.47 & 3.32 & 5.27 \\
\hline
\multicolumn{5}{l}{$^{\mathrm{*}}$SD - Standard Deviation}
\end{tabular}
\label{tab-true_pos}
\end{center}
\end{table}

In further analysis of the prediction error, the false negatives are replaced with value estimates, so each ground truth value has an associated prediction for both methods. To fill in values for false negatives, a simple strategy is applied. If a missing value is directly between two other predictions, the false negative is estimated as the average of the two adjacent predictions. For other missing values, the false negative is estimated as the nearest prediction whether it occurred before or after that time interval. After performing this operation, each method now has equivalent size of paired samples of $654$, $659$, and $657$ for heart rate, diastolic BP, and systolic BP symbols, respectively. Table \ref{tab-false_neg} shows the updated prediction results after estimating false negative values and applying the bias conversions. The mean errors are now all close to zero, but the two methods differ in variance. 

\begin{table}[htbp]
\caption{Error Results with False Negatives Estimated}
\begin{center}
\begin{tabular}{|c|c|c|c|c|}
\hline
\textbf{Method} & \textbf{Statistic} & \textbf{\textit{Heart Rate}}& \textbf{\textit{Diastolic BP}}& \textbf{\textit{Systolic BP}} \\
\hline
\multirow{2}{8mm}{U-Net} & Mean Error & -0.52 & -0.15 & 0.04 \\
\cline{2-5}
& SD* & 2.59 & 3.25 & 2.05 \\
\hline
\multirow{2}{8mm}{TM} & Mean Error & 0.63 & -0.08 & 0.06 \\
\cline{2-5}
& SD* & 4.77 & 3.68 & 6.08 \\
\hline
\multicolumn{5}{l}{$^{\mathrm{*}}$SD - Standard Deviation}
\end{tabular}
\label{tab-false_neg}
\end{center}
\end{table}

\begin{figure*}[ht]
\center
\includegraphics[width=0.8\textwidth]{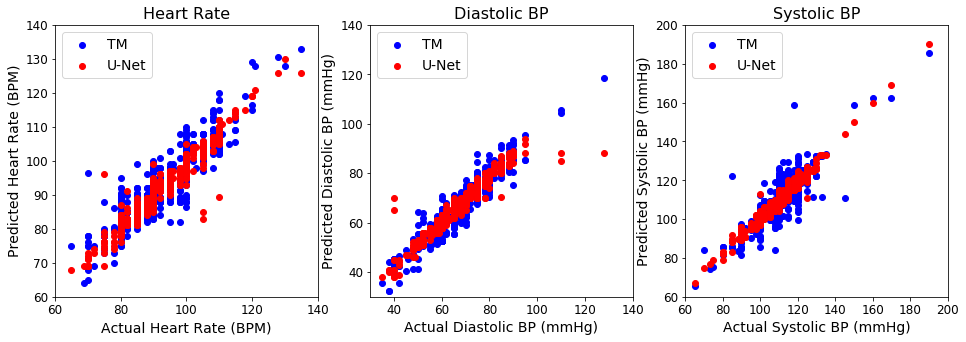} 
\caption{Comparison of Methods with Scatterplots of Predicted vs. Actual Values}
\label{fig-scatters}
\end{figure*}

The differences in variance can also be analyzed with statistical testing. The $F$-test for equality of two variances is applied to determine if the TM variance is significantly higher than the U-Net variance. The null hypothesis of this test is $H_0: \sigma^2_{T} = \sigma^2_{U}$ where $\sigma^2_{T}$ is the variance of the TM error and $\sigma^2_{U}$ is the variance of the U-Net error. The alternative hypothesis is an upper-tailed test of $H_A: \sigma^2_{T} > \sigma^2_{U}$ to test if TM error variance is significantly greater than U-Net. The $F$-statistic for this test is $S^2_{T}/S^2_{U}$ where $S^2_{T}$ is the sample variance for TM error and $S^2_{U}$ is the sample variance for U-Net error. The test statistic is compared to the appropriate $F$-critical value for a 95\% confidence to determine the result of the hypothesis test. Table \ref{tab-var_Ftest} shows the results for the three $F$-tests performed. The null hypothesis is rejected for each symbol, so the variance in U-Net errors is significantly lower than TM. The variances for diastolic BP were close, but due to the large sample size of the data, the difference is still significant.

\begin{table}[htbp]

\caption{Variance Error Statistical Comparison}
\begin{center}
\begin{tabular}{|c|c|c|c|}
\hline
\textbf{$F$-test parameters} & \textbf{\textit{Heart Rate}}& \textbf{\textit{Diastolic BP}}& \textbf{\textit{Systolic BP}} \\
\hline
TM Variance & 22.79 & 13.57 & 37.00 \\
\hline
U-Net Variance & 6.70 & 10.54 & 4.19 \\
\hline
$F$-statistic & 3.40 & 1.29 & 8.83 \\
\hline
$p$-value & $2.4 \times 10^{-52}$ & $6.0 \times 10^{-4}$ & $9.2 \times 10^{-146}$ \\
\hline
Result ($\alpha = 0.05$) & Reject $H_0$ & Reject $H_0$ & Reject $H_0$ \\
\hline
\end{tabular}
\label{tab-var_Ftest}
\end{center}
\end{table}

The predicted versus actual numerical comparison is similar to a regression problem, so the same performance metrics should apply. We applied traditional regression metrics such Mean Squared Error (MSE), Mean Absolute Error (MAE), and $R^2$ to compare the prediction accuracy of U-Net and TM. Table \ref{tab-regression} shows the results of the regression performance metric comparison between U-Net and TM. U-Net outperforms TM for each metric on all three symbols. 

\begin{table}[htbp]
\caption{Regression Performance Metric Comparison}
\begin{center}
\begin{tabular}{|c|c|c|c|c|}
\hline
\textbf{Method} & \textbf{Metric} & \textbf{\textit{Heart Rate}}& \textbf{\textit{Diastolic BP}}& \textbf{\textit{Systolic BP}} \\
\hline
\multirow{3}{8mm}{U-Net} & MSE & 6.97 & 10.48 & 4.19 \\
\cline{2-5}
& MAE & 1.65 & 1.67 & 1.45 \\
\cline{2-5}
& $R^2$ & 0.933 & 0.919 & 0.967 \\
\hline
\multirow{3}{8mm}{TM} & MSE & 23.15 & 13.45 & 36.94 \\
\cline{2-5}
& MAE & 3.89 & 2.69 & 4.17 \\
\cline{2-5}
& $R^2$ & 0.777 & 0.889 & 0.743 \\
\hline
\end{tabular}
\label{tab-regression}
\end{center}
\end{table}

The diastolic BP comparison showed much closer variances and regression metrics than the other two symbols. To investigate this further, Fig. \ref{fig-scatters} shows three separate scatterplots for predicted vs. actual values. On all three graphs, it is apparent that the U-Net results have a tighter and less dispersed fit. For the diastolic BP, it appears that U-Net performance metrics were greatly affected by about five outliers. These outliers are likely due to inaccurate false negative estimates. One limitation with U-Nets is not detecting symbols when the value changes drastically over the 5-min interval. This issue is not widespread, but there were enough to affect the diastolic BP performance comparison. These errors can be alleviated in the future by increasing the size of the training set.

\section{Conclusions and Future Work}

Fast and efficient digitization of paper flowsheets used to collect perioperative data in low and middle income countries is essential to understanding the causes of adverse surgical outcomes in those countries. This paper presents a deep image segmentation approach to achieve this goal through digitizing the values of hand-drawn symbols in the surgical flowsheet graphs. Template matching, an image processing approach, did not perform as well due to an inability to generalize and recognize many different variations in how symbols are drawn. With only 29 annotated images, a deep learning approach would not normally seem plausible. In this dataset, the images all have the same setting of a white background with black gridlines. The objects being detected are simple circles or angled arrows. The standardized background and simple objects were likely big factors in enabling the success of a deep learning approach on such a small labeled dataset. Although there were only 29 images, each of the three symbols had over 650 individual annotations.
    
The medical researchers interested in the follow-on study of the perioperative data desire a predictive error of $\pm5$ for all three symbols. The U-Net approach achieves that goal as the three standard deviations for true positive error were all less than $2.5$. Assuming a normal distribution, this indicates that over 95\% of predicted values are expected to fall within $\pm5$ error. This research can benefit any medical center or country that does not currently have a process of digitizing the data from these surgical flowsheets.

Future work in this area should begin with improving the existing methodology. The U-Net segmentation accuracy can be improved by annotating more graphs for the training set. The post-processing methods can be adapted and increased in rigor to more accurately capture values from the segmentation masks. The graphs are very rectangular, so a zero-padding approach was used to process the images through the Residual U-Net. An approach that patches the graph into smaller images will enable the model to increase in complexity and fit within memory constraints, but can be more difficult to train. There are other semantic segmentation methods or deep learning computer vision approaches, such as object recognition CNNs, that may also perform well at digitizing the graph data.

\bibliographystyle{IEEEtran}
\bibliography{main}

\end{document}